\documentclass[aps,prb,twocolumn, showpacs]{revtex4}
\usepackage{amsmath,amssymb}
\usepackage{graphicx}

\begin{document}

\title{Metallic monoclinic phase in VO$_2$ induced by electrochemical gating: in-situ Raman study}
\author{Satyendra Nath Gupta, Anand Pal, D. V. S. Muthu, P. S. Anil Kumar and A. K. Sood}
\thanks{To whom correspondence should be addressed}
\email{asood@physics.iisc.ernet.in}
\affiliation{Department of Physics, Indian Institute of Science, Bangalore-560012,India}

\date{\today}

\begin{abstract}
We report in-situ Raman scattering studies of electrochemically top gated VO$_2$ thin film to address metal-insulator transition (MIT) under gating. The room temperature monoclinic insulating phase goes to metallic state at a gate voltage of 2.6 V. However, the number of Raman modes do not change with electrolyte gating showing that the metallic phase is still monoclinic. The high frequency Raman mode A$_g$(7) near 616 cm$^{-1}$ ascribed to V-O vibration of bond length 2.06 \AA~ in VO$_6$ octahedra hardens with increasing gate voltage and the B$_g$(3) mode near 654 cm$^{-1}$ softens. This shows that the distortion of the VO$_6$ octahedra in the monoclinic phase decreases with gating.
The time dependent Raman data  at  fixed gate voltages of 1 V (for 50 minute, showing enhancement of conductivity by a factor of 50) and 2 V (for 130 minute, showing further increase in conductivity by a factor of 5) show similar changes in high frequency Raman modes A$_g$(7) and B$_g$(3) as observed in gating.  This slow change in conductance together with  Raman frequency changes  show that the governing mechanism for metalization is more likely to the diffusion controlled oxygen vacancy formation due to the applied electric field.
\end{abstract}

\maketitle
The metal-insulator based transistor is a most promising candidate for surpassing  limits of classical electrical charge based device with the potential of higher speed and low power consumptions \cite{pershin2011memory,yang2011oxide,aetukuri2013control}. 
Vanadium dioxide (VO$_2$) is a well-known strongly correlated material that undergoes metal insulator-transition (MIT) with five orders of magnitude change in resistance near  T$_{MI}$ $\approx$ 340 K \cite{PhysRevLett.3.34,PhysRev.185.1022}. This thermally driven  MIT  is  accompanied  by structural  phase transition (SPT), from the high temperature metallic rutile (R)  structure (P4$_2$/mnm) to  insulating  ( band gap $\sim$ 0.6 eV) with  monoclinic (M) structure (P2$_1$/c)  ~\cite{goodenough1971two,mott1968metal}. The MIT of VO$_2$ can be controlled not only by temperature but also by light and electrical current \cite{PhysRevLett.3.34,nakano2012collective,morrison2014photoinduced}. This unique characteristic  makes VO$_2$ a compelling candidate for microelectronic devices, terahertz devices, energy harvesting systems etc  ~\cite{wu2013design,yang2011oxide,liu2012terahertz,cao2009thermoelectric}. Controlling and tailoring the conductivity  of VO$_2$ by  electric field  have a potential for realizing next-generation hybrid multifunctional, low power logic and non-volatile memory devices\cite{sengupta2011field}. This type of switching process requires  high electric field, which cannot be achieved by present day dielectric gates and hence calls for electrolytic top gating   \cite{asanuma2010tuning,scherwitzl2010electric,hatano2013gate} as has been done using ionic liquid (IL) by two groups of Iwasa et al.~\cite{nakano2012collective,okuyama2014gate} and Jeong et al.~\cite{jeong2013suppression,jeong2015giant}.  Both the groups showed that above a certain positive gate voltage, the MIT temperature is suppressed.  However,  different mechanisms have been proposed  for electric field induced metalization  in VO$_2$ films. Iwasa  el.~\cite{nakano2012collective,okuyama2014gate} suggested that the three dimensional metallic ground state  emerges due to the collective bulk delocalization of electrons driven by electrostatic charging at the surface of  VO$_2$. On the other hand, Jeong et al.~\cite{jeong2013suppression,jeong2015giant} proposed that the metalization of  VO$_2$ is due to the electric field induced oxygen vacancy formation, arising from the migration  of oxygen ions from the  channels in to the ionic liquid. Further Iwasa et al.~\cite{nakano2012collective,okuyama2014gate} claimed that the MIT was accompanied by an increase of the unit cell along the $c$-axis  by 1\% on gating. This metallic state with larger $c$-axis lattice parameter is completely reversible and different from the monoclinic phase.  On the other hand, Jeong et al.~\cite{jeong2013suppression,jeong2015giant} showed that above the critical gate voltage, VO$_2$ turns metallic, remain in monoclinic phase, and the $c$-axis expands more than  3\%. This metallic state and the associated structural distortion remain even if the gate voltage is removed. However, initial monoclinic undistorted insulating state can be recovered by applying reverse gate voltage. Hence the intriguing question about the role of SPT  and the underlying physical mechanism of electric field induced metalization process of VO$_2$  is still open.  This  motivated us to do in-situ Raman scattering experiments on devices as a function of electrolytic gating. Raman scattering has proven to be a powerful  probe for the local site symmetry,  structural changes and electron phonon coupling ~\cite{lee2010anomalous,berkdemir2013identification, PhysRevLett.98.166802, das2008monitoring,PhysRevB.85.161403,chakraborty2016electron}.  Our study shows that the   VO$_2$ insulating monoclinic phase becomes metallic  at a gate voltage of  2.6 V  retaining the monoclinic symmetry. Further, our data  suggest that the mechanism of metalization  is likely due to oxygen vacancy formation as shown by Jeong et al.~\cite{jeong2013suppression,jeong2015giant}.

\begin{figure}
\begin{center}
\includegraphics [width=0.45\textwidth]{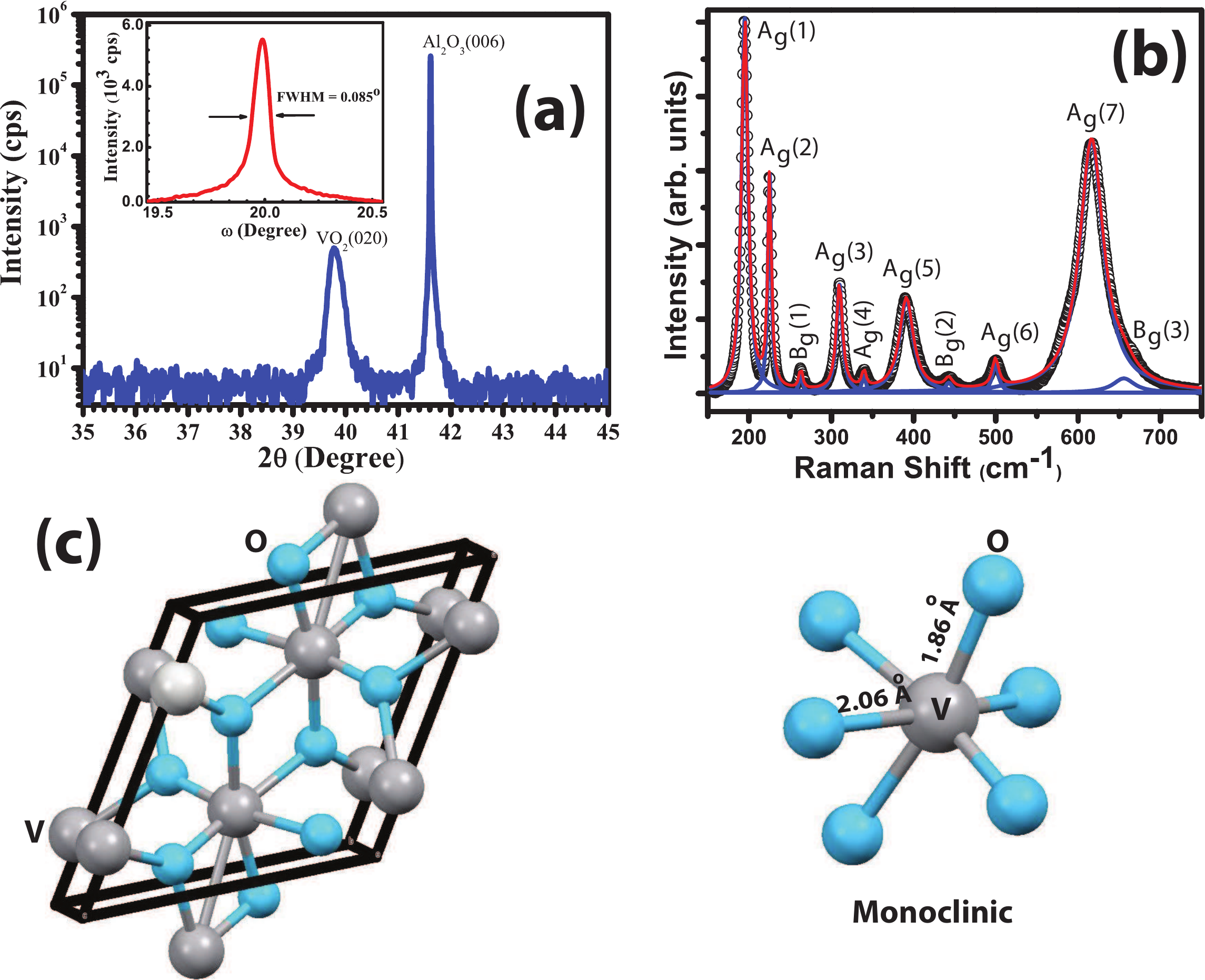}
\end{center}
\caption{\label{xrd raman structure} (Color Online) (a) XRD pattern of VO$_2$ film at room temperature. Inset: XRD rocking curve of VO$_2$ (020) diffraction peak. (b) Raman spectra of VO$_2$ film at room temperature. (c) Crystal structure of VO$_2$ film in the Monoclinic  phases along with its  VO$_6$ octahedra with V-O bond lengths.} 

\end{figure}

\begin{figure}
\begin{center}
\includegraphics [width=0.45\textwidth]{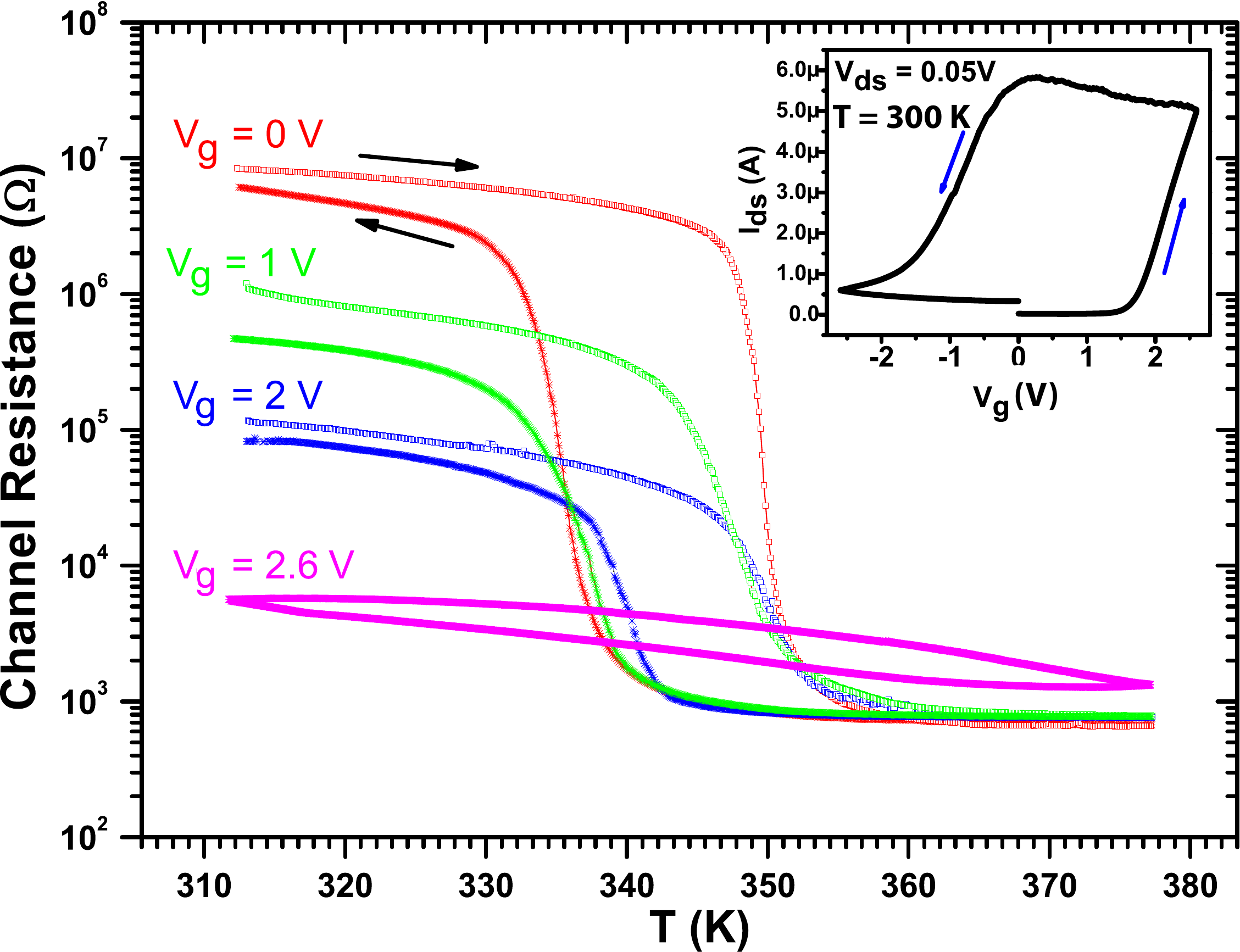}
\end{center}
\caption{\label{RT with gate} (Color Online) Suppression of MIT in VO$_2$ film. Temperature dependence of VO$_2$ Chanel resistance at different gate voltages varying from 0 to 2.6 V. Inset: hysteresis in the channel conductance, where I$_{ds}$, V$_{ds}$ and V$_g$ are source drain current, source drain voltage and gate voltage respectively} 

\end{figure}

The epitaxial VO$_2$ thin films of thickness 50~nm were grown on $c$-plane Al$_2$O$_3$ substrate by  Pulsed Laser deposition (PLD) using V$_2$O$_5$ as the target at 580~$^o$C  under 10 mtorr oxygen pressure. The KrF Excimer laser with laser fluence of ~1.5 J/cm$^2$ and 5 Hz repetition rate was used to ablate the rotating V$_2$O$_5$ target. The target was made by compaction of commercially available V$_2$O$_5$ powder and sintered at 630~$^o$C for 24 hours.  The multiple oxidation state of vanadium makes it  very sensitive to partial oxygen pressure. We found that the optimum oxygen pressure window is between 8 to 15 mtorr for oriented growth of VO$_2$. Before thin-film deposition,  Al$_2$O$_3$ substrate (0001)  was ultrasonically pre-cleaned by  trichloroethylene, acetone and isopropyl alcohol (IPA) for 10 min each and baked at 580~$^o$C for a hour under 1 $\mu$torr in the deposition chamber. The as grown VO$_2$ films were characterized by X-ray diffraction measurements  using Rigaku SmartLab x-ray Diffractometer with monochromatic Cu- K$\alpha$ radiation.  Fig.\ref{xrd raman structure} (a) shows the room temperature XRD diffraction pattern of the film, which indicates that the VO$_2$ film deposited on the c-plane Al$_2$O$_3$ is in single phase and oriented in (020) direction.  The XRD rocking curve of VO$_2$ [020] diffraction peak is shown in the inset of  Fig.\ref{xrd raman structure}(a). The relatively small value of full width at half maximum (FWHM)  (0.09$^0$) of rocking curve reflects  high quality of   VO$_2$ films on $c$-plane Al$_2$O$_3$. The slight deviation from an ideal Gaussian shape of the (020) peak  and corresponding rocking curve  indicate the residual strain in the thin film due to the lattice mismatch between the film and substrate. The as grown VO$_2$ film shows MIT near 340K with a change in resistance of $\sim$10$^4$ (Fig.\ref{RT with gate}). 

The VO$_2$ thin film was structured in a channel area of $200 \times 20$ $\mu$m$^2$ using photolithography followed by reactive chlorine ion etching. A hard-baked photoresist was used as an etching mask. The remaining parts of the VO$_2$ films were removed by reactive ion beam etching in a commercial reactive ion etcher in the Ar (24 SCCM flow)/Cl$_2$ (8 SCCM) gas environment at 150 Watt  power at 15~ mtorr pressure for 90 sec. Source, drain and side gate electrodes were fabricated by photolithography followed by sputter deposition of 10/100 nm of Cr/Au. The  separation  between gate electrodes and the channel was $\approx$ 50 $\mu$m.

Confocal Raman  measurements were carried out in backscattering geometry using Labram HR-800, coupled  with a Peltier cooled CCD, using laser wavelength of 532 nm and a long working distance 50X objective with numerical aperture 0.45. For electrical measurements, drain-source and gate voltages were applied from Keithley 2400 source meters which also measures the respective currents. A ~200~nl of 1-ethyl-3-methylimidazolium bis(trifluoromethylsulfonyl)-imide  was drop-casted on the device in such way that the droplet covered the whole Chanel and sufficient part of the lateral gate electrode. 

Group theory predicts 18 Raman active modes (9A$_g$ and 9 B$_g$) for the low temperature monoclinic phase of VO$_2$. The experimentally observed Raman modes are  shown in Fig.\ref{xrd raman structure} (b). Experimental data (shown by open circles) are fitted with a sum of ten Lorentzians (shown by lines)  The mode assignment has been done  following previously reported Raman data ~\cite{schilbe2002raman,schilbe2004lattice}. The crystal structure of VO$_2$ film below   T$_{MI}$ is shown in Fig.\ref{xrd raman structure} (c).  The VO$_6$ octahedra of the monoclinic phase  has two types of  V-O bonds of bond lengths 2.06 \AA~ and 1.86 \AA. Jeong et al.~\cite{jeong2015giant}  showed using x-ray diffraction that with IL gating on the monoclinic phase, both the bond lengths change to 1.95 \AA.
Comparing the phonon frequencies of VO$_2$ with isostructural NbO$_2$, which exhibits a  similar Raman spectrum~\cite{zhao2004optical} as monoclinic VO$_2$,  A$_g$(1) and A$_g$(2) frequencies scale with the mass of the transition metal  while the frequency of the A$_g$(7) and B$_g$(3) mode  scale with the reduced mass between the oxygen and the transition metal. Thus the lowest frequency phonons A$_g$(1) and A$_g$(2) are ascribed to the V-ions motion within the V-V chains and the high frequency modes at $\sim$ 616 $cm^{-1}$ ( A$_g$(7)) and $\sim$ 654 $cm^{-1}$  (B$_g$(3)) are ascribed to  V-O modes due to the large difference between the mass of V and O atoms. In this study, we have focused on the  A$_g$(1),  A$_g$(2), A$_g$(7) and B$_g$(3) modes, since they give clear information about the V-V and V-O bonds.

\begin{figure}
\begin{center}
\includegraphics [width=0.45\textwidth]{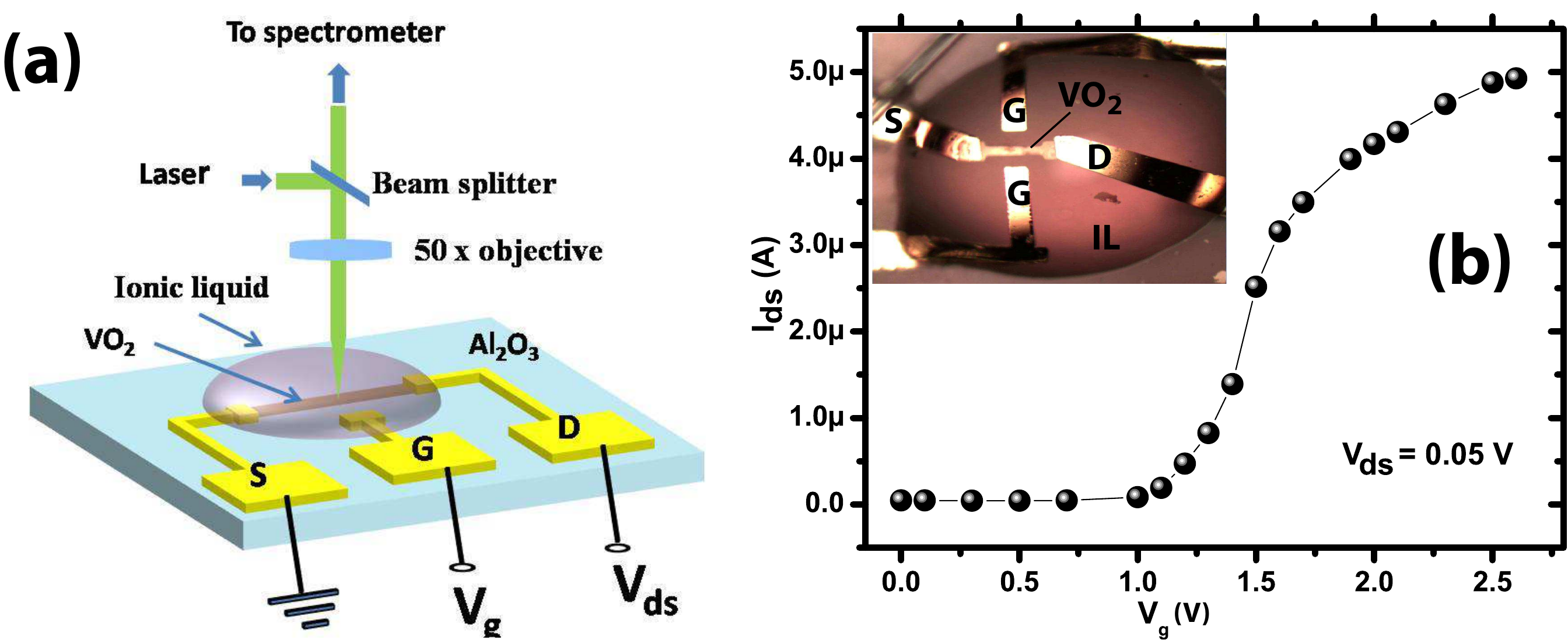}
\end{center}
\caption{\label{Gate characteristic} (Color Online) (a) Schematic of our experimental set up. (b) Gate characteristic for the  device. Inset: optical image of the device showing a droplet of IL.}
\end{figure}

\begin{figure}
\begin{center}
\includegraphics [width=0.45\textwidth]{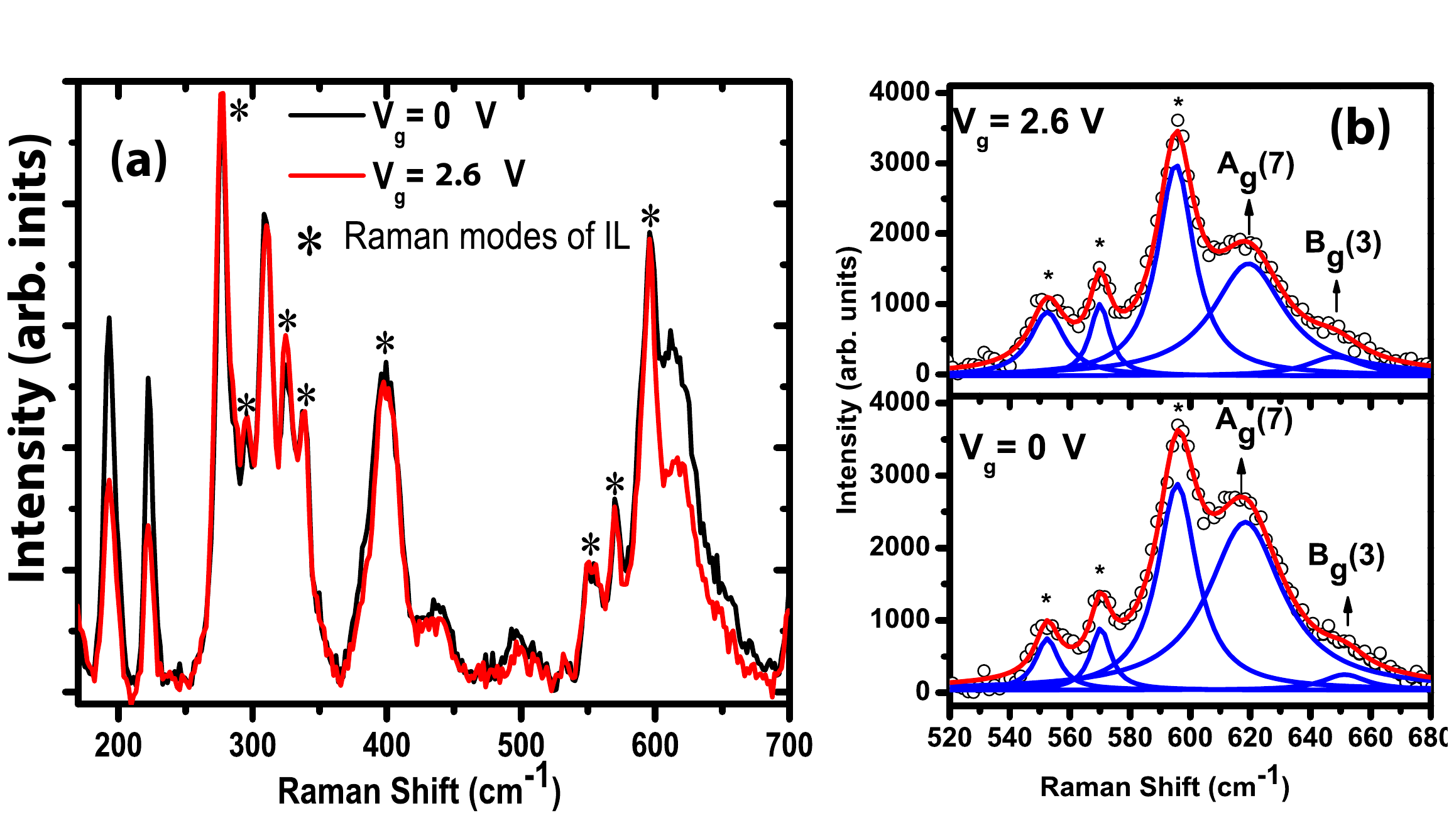}
\end{center}
\caption{\label{Raman spectra} (Color Online) (a) Raman spectra of VO$_2$ film at 0 and 2.6 gate voltages. (b) Magnified  Raman spectra between 520 to 680 $cm^{-1}$. Open circles are experimental data and solid lines are Lorentzian  fitting. The peaks marked with * are due to IL.}
\end{figure}

\begin{figure}
\begin{center}
\includegraphics [width=0.45\textwidth]{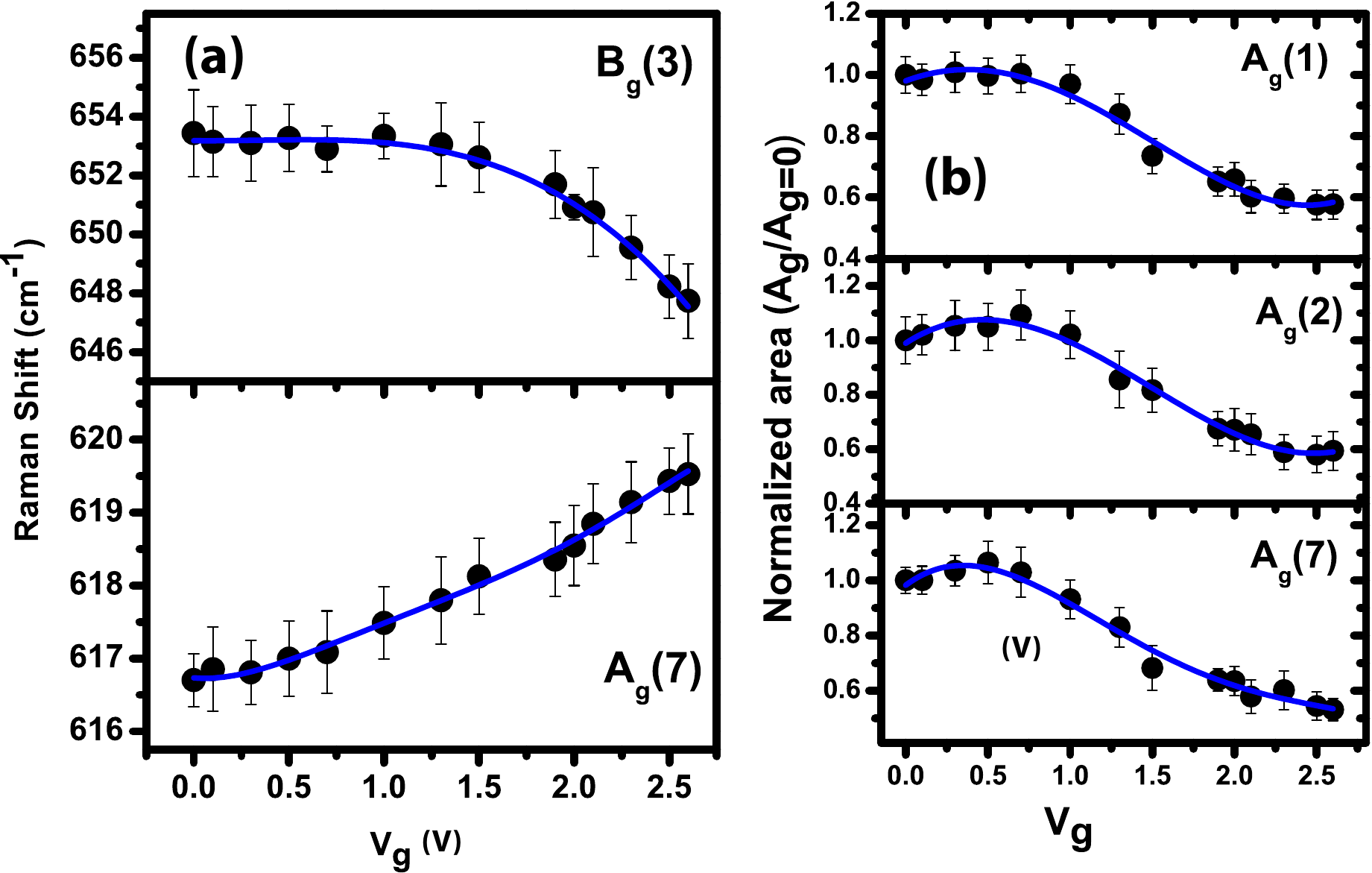}
\end{center}
\caption{\label{phonon frequency} (Color Online)(a) Dependence of  phonon frequencies  of A$_g$(7) and B$_g$(3) modes with gate voltages. (b) Dependence of  Area of  A$_g$(1),  A$_g$(2) and A$_g$(7) modes with gate voltages. The solid blue lines are guide to the eye.}
\end{figure}

\begin{figure}
\begin{center}
\includegraphics [width=0.45\textwidth]{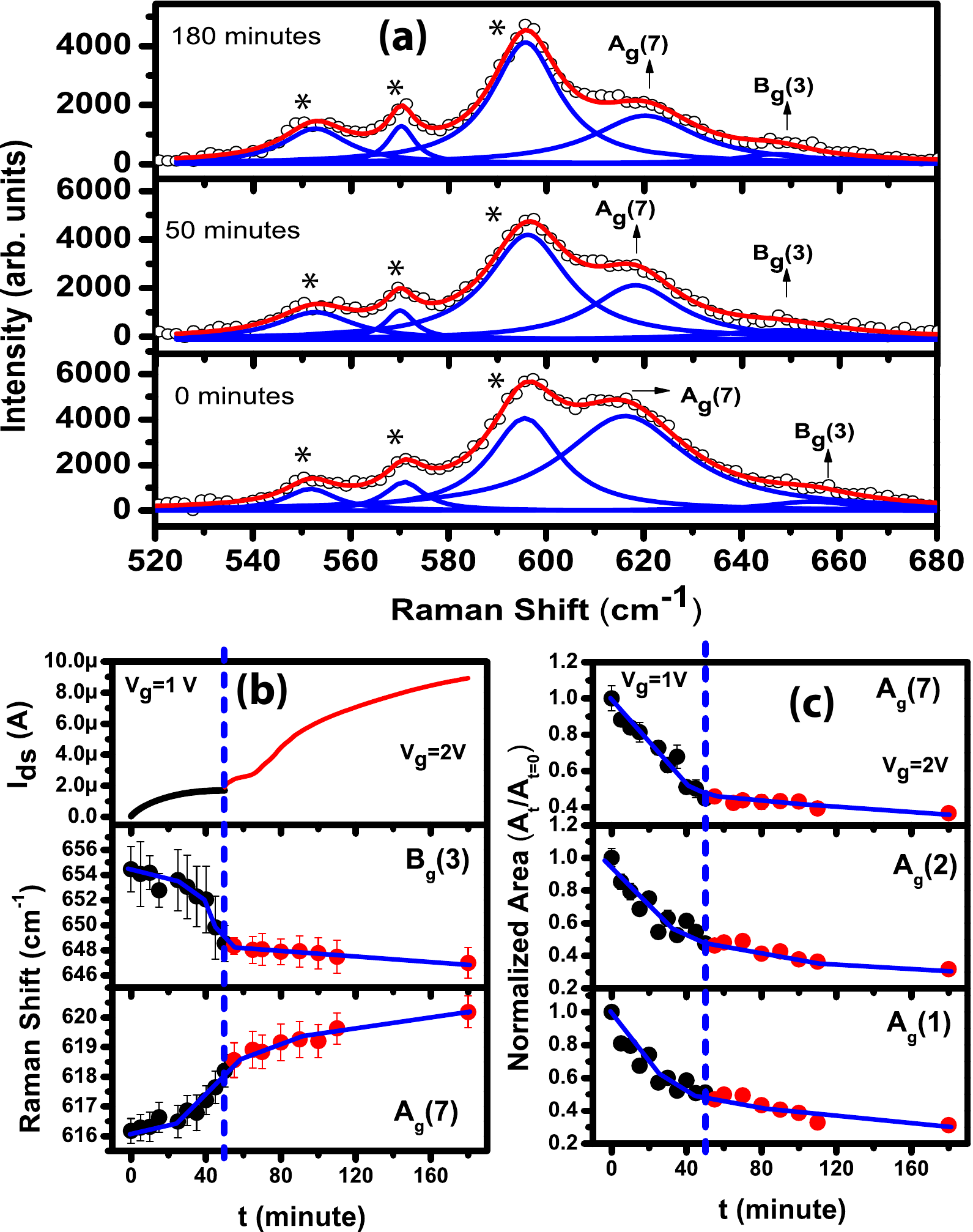}
\end{center}
\caption{\label{time data} (Color Online)(a) Time dependence Raman spectra. Solid blue lines are individual Lorentzian peak fitting while solid red lines are sum of all fittings. The peaks marked with * are due to IL. (b) Time dependence of current and phonon frequency  of A$_g$(7) and  B$_g$(3) modes. (c) Time dependence of Area of A$_g$(1), A$_g$(2) and A$_g$(7) modes. The solid blue lines are guide to the eye. The vertical dashed lines show two different region of gate voltage.}
\end{figure}

Fig.\ref{RT with gate}  shows the temperature dependent channel resistance  at different gate voltages (V$_g$) applied in the insulating state at 300 K. The Channel resistance decrease as the gate bias is increased which result in reduction of MIT. At a gate voltage of  2.6 V, complete suppression of MIT is observed indicating the emergence of the metallic ground state. Our data are in good agreement with the previously reported data\cite{jeong2013suppression,nakano2012collective} .  The inset of Fig.\ref{RT with gate} shows the hysteresis in the channel conductance centered about V$_g$ = 0 V (I$_{ds}$ vs V$_g$ where I$_{ds}$, V$_{ds}$ and V$_g$ are source-drain current, source-drain voltage and gate voltage, respectively) recorded at 300 K. The gate voltage was swept by 0.01 V/second.  Once the metallic state is achieved, it remains in the high conductance state even if the gate is removed but insulating state can be  recovered by applying reverse gate voltage.

In order to understand whether the metallic ground state achieved by applying 2.6 V gate voltage is similar to the high temperature tetragonal structure or not, we have done in-situ   Raman study at different gate voltages. The schematic of experimental setup for in-situ   Raman measurements is shown in Fig.\ref{Gate characteristic} (a). The device transfer characteristic as shown in Fig.\ref{Gate characteristic} (b) shows that the overall conductance of the VO$_2$ film increases by nearly 100 times at a gate voltage of 2.6 V. The optical image of the device along with a droplet of the IL is shown in the inset of Fig.\ref{Gate characteristic} (b).  At each gate voltage, we waited for 5 minutes and then  Raman spectrum was recorded for 2 minutes.
Raman spectra recorded at 0 V and 2.6 V gate voltages are compared in  Fig.\ref{Raman spectra} (a). The peaks marked by asterix are due to Raman lines of ionic liquid. The low frequency phonons A$_g$(1) and A$_g$(2) assigned to V-V modes  do not show any change in Raman frequency  with the electrolyte gating. As A$_g$(1) and A$_g$(2) phonons are related to the in plane V-V vibrations \cite{PhysRevLett.98.196406}, this indicates that the structure in the $ab$-plane is unaffected, which is in good agreement with in-situ synchrotron x-ray diffraction and absorption experiments on IL gated VO$_2$ film~\cite{jeong2015giant}.
The high frequency phonons A$_g$(7) and B$_g$(3)  show  changes as shown in   Fig.\ref{Raman spectra} (b).  In order to estimate the phonon frequencies,  FWHM and area of the phonon modes at all investigated gate voltages, Lorentzian line shapes are  fitted to the Raman spectra.  Fig.\ref{Raman spectra} (b) shows the  fitted  Raman spectra collected at  gate voltage of 0 and 2.6 V.

The evolution of Raman frequencies of  A$_g$(7) and B$_g$(3) modes with gate voltage is shown in Fig.\ref{phonon frequency} (a). It clearly reveals that the A$_g$(7) mode hardens  while the B$_g$(3) mode softens with increasing gate voltage. The hardening  (softening) of phonon frequency indicates that V-O bond lengths are decreasing  (increasing) with the increasing gate voltage.  Thus on gating, the V-O  bond length corresponding to  A$_g$(7) mode decreases and bond length associated with the  B$_g$(3) mode increases. 
Jeong et al.~\cite{jeong2015giant} showed from in-situ synchrotron x-ray diffraction and absorption experiments that there are two types of V-O  bonds of bond length 2.06 \AA~ and 1.86 \AA~  in VO$_6$ octahedra in the monoclinic phase and both the bond lengths  converge to 1.95 \AA~ on electrolyte gating. Thus one of the bond is contracting and the other bond is expanding on electrolyte gating. Keeping this in mind, we assign A$_g$(7) mode to the vibration of V-O  bond of bond length 2.06 \AA~ and the  B$_g$(3) mode may be due to vibration of   V-O bond of bond length 1.86 \AA.  Our Raman data on A$_g$(7) and B$_g$(3) suggest that the symmetry of VO$_6$ octahedra is increasing with electrolyte gating, although not as much as seen in x-ray absorption experiments\cite{jeong2015giant}.   The area of the A$_g$(7), A$_g$(1) and  A$_g$(2) modes decrease by $\sim 40\%$  with gating as shown in  Fig.\ref{phonon frequency} (b). This decrease in the Raman intensity can be due to  the field-induced  conductivity  which attenuates the excitation beam and increases optical losses for the Raman signal.

In order to address whether the mechanism for gate induced MIT  is electronic or oxygen vacancy formation, we measured the kinetics of gate voltage induced changes. The kinetics at a given gate voltage will be much slower for the oxygen vacancy-formation mechanism as compared to the electrostatic charges. We first applied 1 V of gate voltage and recorded Raman spectra and current as a function  of time for 50 minutes. Then  the gate voltage was changed to 2 V and  Raman spectrum and current were recorded  till 130 minutes. At a gate voltage of 1 V, VO$_2$ is in the  insulating state at room temperature as evident from Fig.\ref{Gate characteristic} (b).  The Raman spectra recorded at three different times are shown in  Fig.\ref{time data} (a) in the spectral range 520 to 680 cm$^{-1}$. A$_g$(7) and  B$_g$(3) modes show changes in Raman frequency with time. A$_g$(1) and  A$_g$(2) modes do not show any change in their peak position with time. Fig.\ref{time data} (b) shows that current increases slowly with time and shows a tendency to saturate beyond 50 minutes (black line). After 50 minutes gate voltage was changed to 2 V and then current started to increase again with time (red line). The change in current is by a factor of five in 130 minutes, giving a total change of current from the pristine value by a factor of 250 in 180 minutes. Another important aspect to note is that at the gate voltage of 2 V, the current did not show any tendency to saturate till 130 minutes.  If the mechanism was electronic, it would  have much faster response ($<$ a few seconds). Since the field induced removal of oxygen is diffusion controlled, it can take a long time to saturate the change in conductance . Raman spectra recorded in the in-situ condition at different times follow the slow change in conductance. The hardening of  phonon frequency  of  A$_g$(7) mode (black circle for V$_g$=1 V and red circle for V$_g$=2 V)  with time show that the   V-O bond corresponding to this mode contracts    with time at a fixed gate voltage. The changes in B$_g$(3) mode is also similar to gating data. The phonon frequency  of  A$_g$(1) and  A$_g$(2) modes  do not change with time, consistent with the gate voltage data. The area of A$_g$(1),   A$_g$(2) and A$_g$(7) also change with time as shown in Fig.\ref{time data} (c) whereas the area of B$_g$(3) mode does not change significantly. It is also evident from the above figures that the change in phonon frequency of A$_g$(7) and  B$_g$(3) for V$_g$=1 V (current enhancement factor of 50) is large compared to V$_g$=2 V (current enhancement factor of 5). Same is also seen for the area of A$_g$(1), A$_g$(2) and A$_g$(7) modes.

In conclusion, we have shown that the room temperature monoclinic insulating  phase is driven into the metallic state with IL gating but retains its monoclinic symmetry. Our gate voltage dependent Raman results show that the  IL gating reduces the distortion of VO$_6$ octahedra while the structure in the $ab$-plane remains unaffected. Further, the slow kinetics of gate voltage induced changes suggest that the mechanism for electric field induced   metalization in VO$_2$ film is diffusion controlled oxygen vacancy formation. It would be very interesting to understand theoretically the large change in conductivity, keeping the monoclinic phase with small change in the VO$_6$ octahedra.

AKS and PSAK thanks the Department of Science and Technology, India   for financial support. A.P. acknowledges financial support  from Dr. D.S.Kothari postdoctoral fellowship (UGC-DSKPDF) program.

\end{document}